\documentclass{revtex4}
\usepackage{times}
\usepackage{epsfig,amssymb,amsfonts,amsmath}
\usepackage[usenames]{color}

\usepackage{amssymb}
\usepackage{amsmath}
\usepackage{amsfonts}
\usepackage{graphicx}
\usepackage{ulem} 
\usepackage{cancel}
\usepackage{extarrows}
\providecommand{\rbr}[1]{\left( #1 \right)}%

\begin{document}

\title[ ]{Route from discreteness to the continuum for the Tsallis $q$-entropy}
\author{$^{1}$Thomas Oikonomou}
\email{thomas.oikonomou@nu.edu.kz}
\author{$^{2}$G. Baris Bagci}
\email{gbb0002@hotmail.com}

\affiliation{$^{1}$Department of Physics, School of Science and Technology, Nazarbayev University, Astana 010000, Kazakhstan}
\affiliation{$^{2}$Department of Materials Science and Nanotechnology Engineering, TOBB University of Economics and Technology, 06560 Ankara, Turkey}
%
%
%

\begin{abstract}
The existence and exact form of the continuum expression of the discrete nonlogarithmic $q$-entropy is an important open problem in generalized thermostatistics, since its possible lack implies that nonlogarithmic $q$-entropy is irrelevant for the continuous classical systems. In this work, we show how the discrete nonlogarithmic $q$-entropy in fact converges in the continuous limit and the negative of the $q$-entropy with continuous variables is demonstrated to lead to the (Csisz{\'a}r type) $q$-relative entropy just as the relation between the continuous Boltzmann-Gibbs expression and the Kullback-Leibler relative entropy. As a result, we conclude that there is no obstacle for the applicability of the $q$-entropy to the continuous classical physical systems.

\end{abstract}

\eid{ }
\date{\today }
\startpage{1}
\endpage{1}
\maketitle


Since its advent, the nonadditive $q$-entropy \cite{Tsallis88,Tsallisbook} has found numerous fields of application in many diverse fields \cite{Chang,Tribeche,Rastegin,bath,Portesi,Campisi,Rotundo,Van1,high2,high3,third}. 
Despite this apparent progress in the field, however, there have been some criticisms regarding its applicability and scope. Among such criticisms, one can particularly cite the ones related to the Bayesian updating procedure \cite{Presse}, Lesche stability \cite{Lesche,Abe1,Lutsko}, and the methodology of the entropy maximization \cite{OikBagci2010}.

Recently, Abe pinpointed that the nonadditive $q$-entropy is inherently limited to the finite discrete systems, since its continuum expression has not been obtained yet \cite{Abecomment} (see also Refs. \cite{Andresen,Abereply}). In this work, we show that one can indeed obtain the concomitant continuum expressions of the nonadditive entropy and therefore point out that the nonadditive $q$-entropy can also be used for continuous physical systems.

Before proceeding further with the nonadditive case, one should be convinced why taking the route from discreteness to a continuum is essential concerning any entropy measure in general. Setting the Boltzmann constant to unity, the finite discrete Boltzmann-Gibbs (BG) entropy reads
\begin{equation} \label{BG}
S(\{p\})=\sum\limits_{i=1}^{n}p_{i}\ln (1/p_{i})
\end{equation}
where $p_i$ denotes the probability of the $i$th event. Let us now consider its continuous counterpart to be the following expression 
\begin{equation}\label{BGint}
S(\rho)=\int\limits_{a}^{b}\rho(x)\ln \left(\frac{1}{\rho(x)}\right) \mathrm{d}x
\end{equation}
where $\rho(x)$ is a probability density function satisfying the normalization condition in the interval $[a,b]$.

Although the continuous expression above seems reasonable at first sight, it has three serious drawbacks. First, the continuous version in Eq. (\ref{BGint}) has an overall unit of $\log($length$)$ whereas the discrete entropy in Eq. (\ref{BG}) is dimensionless \cite{Uffink}. Second, the probability density $S(\rho)$ is not invariant with respect to coordinate transformations \cite{Uffink}. Last but not the least, the discrete BG entropy $S(\{p\})$ in the $n\rightarrow\infty$ limit and $S(\rho)$ yield different results \cite{Uffink}: To see this more explicitly, consider a uniform distribution $\rho(x)$ in the interval $[a,b]$ as $1/(b-a) $ so that its discrete counterpart $p(x_{i})$ is given by $1/n$ obtained through dividing the same interval $[a,b]$ into $n$ equal subintervals where the index $i$ runs from $1$ to $n$. Then, the continuous entropy $S(\rho)$ for this uniform distribution yields $\ln(b-a)$ while the discrete expression $S(\{p\})$ attains infinity in the $n\rightarrow\infty$ limit. In other words, the continuum version of the discrete entropy does not converge to the value obtained through the continuous version for the uniform distribution. Therefore, the continuous version of the discrete BG entropy $S(\{p\})$ cannot be $S(\rho)$.

The solution of the discrete-to-continuum transition for the BG entropy is already known \cite{Jaynes}. In order to extend BG entropy to the continuum, we assume some discrete points $x_{i}$ with $i=1,2,\ldots,n$ and $x_1<\cdots<x_n$ filling the interval $[a,b]$ so that one has a factorizable discrete probability $p_{i}$ \cite{Jaynes} as 
\begin{equation}\label{interval}
p_{i}= \rho(x_i)\Delta x_i\,,
\qquad\Delta x_i=\frac{1}{n\,m(x_i)}\,
\end{equation}
with the property
\begin{eqnarray}\label{norm}
\sum_{i=1}^{n}\rho(x_i)\Delta x_i=1
\qquad \overset{n\to\infty}{\longrightarrow} \qquad
\int_a^b \rho(x)\mathrm{d}x=1\,.
\end{eqnarray}
Substitution of Eq. (\ref{interval}) into the discrete entropy expression given by Eq. (\ref{BG}) yields
\begin{equation}\label{BGint1}
S(\{p\})=\sum\limits_{i=1}^{n}p_i 
\ln \left( \frac{m(x_i)}{\rho(x_i) }\right) + \ln(n)\,
\end{equation}
where we have also made use of the normalization $\sum_{i=1}^{n}p_i=1$.
Equation (\ref{BGint1}) can now be rewritten as
\begin{equation}\label{BGint2}
S(\{p\})=\sum\limits_{i=1}^{n}\rho(x_i)
\ln \left( \frac{m(x_i) }{\rho(x_i) }\right) \Delta x_i 
+\ln (n)
\end{equation}
so that the above summation in the $n\rightarrow\infty$ limit  finally yields the following continuous expression
\begin{equation}\label{relent2}
\lim_{n\to\infty} S(\{p\})= S(\rho)=
\int\limits_{a}^{b}\rho (x) 
\ln \left( \frac{m(x) }{\rho(x) }\right)\mathrm{d}x\,,
\end{equation}
where the additive divergent term $\lim_{n\rightarrow\infty} \ln \left( n\right)$ is omitted since the entropy is not absolute, but only its change can be measured \cite{Abecomment}. It is worth remarking that the continuous entropy expression given by Eq. (\ref{relent2}) is dimensionless like  its discrete counterpart and invariant under different reparametrization of continuum.

Note that the usual discrete nonadditive $q$-entropy, i.e., $S_q=\sum_{i=1}^{n}p_i\ln_q(1/p_i)$ \cite{Tsallis88, Tsallisbook}  ($\ln_q(x)$ is defined in Eq. (\ref{q-log})), cannot be adopted, since it does not converge in the continuous limit \cite{Abecomment}. Therefore, we consider
%
%
\begin{equation}\label{q-entropy1}
S_{q}(\{p\})=n^{q-1}\sum\limits_{i=1}^{n}p_{i}\ln
_{q}\left(1/p_i\right)\,,
\end{equation}
where the $q$-logarithm \cite{OikBagci2009} is defined as
\begin{equation}\label{q-log}
\ln_q(x)\equiv\frac{x^{1-q}-1}{1-q}\,,
\end{equation}
which becomes the ordinary logarithm in the $q\rightarrow1$ limit so that the nonadditive entropy becomes the BG entropy. 
The discrete entropy expression in Eq. (\ref{q-entropy1}) has an additional multiplicative term $n^{q-1}$ compared to the usual nonadditive entropy expression \cite{Tsallis88,Tsallisbook}. As we show below, this term is required for convergence and therefore can be called the convergence factor (see Eq. (\ref{KL4}) below for more on its justification).

In order to extend the discrete expression above to the continuum, we consider the same apparatus as before (see Eq. (\ref{interval}) and related explanations above it) with the exception that we now have $\Delta x_i = \frac{1}{n\,m_{q}(x_i)}$. The measure $m_{q}(x_i)$ is the $q$-deformed form of the previous measure $m(x_i)$ in Eq. (\ref{interval}) to account for the nonadditivity as also noted in Ref. \cite{Abecomment} (see Eq. (10) therein). Therefore, the probability normalization condition in Eq. (\ref{norm}) is satisfied in the case of the nonadditive $q$-entropy as well albeit now under $m_q(x_i)$ so that
\begin{equation}\label{q-entropy2}
S_{q}(\{p\})=
n^{q-1}\sum\limits_{i=1}^{n} p_i 
\ln_{q}\left( \frac{nm_q(x_i)}{\rho(x_i)} \right)
=
n^{q-1}\left[\sum\limits_{i=1}^{n} p_i
\ln_{q} \left( \frac{m_q(x_i)}{\rho(x_i)} \right) 
- \ln_{q}\left(1/n\right)\sum_{i=1}^{n}p_i^q\right]\, .
\end{equation}
Note now that using Eqs. (\ref{q-entropy1}) and (\ref{q-log}), the following relation is seen to hold
\begin{equation}\label{relo}
\sum_{i=1}^{n}p_i^q=[1+(1-q)n^{1-q}\,S_q]\, .
\end{equation}
The substitution of the relation above into Eq. (\ref{q-entropy2}) yields the analogous expression of the Shannon entropy in Eq. (\ref{BGint2})
\begin{equation}\label{mq-measure}
S_q(\{p\})=\sum\limits_{i=1}^{n} \rho(x_i)
\ln_{q} \left( \frac{m_q(x_i)}{\rho(x_i)} \right) \Delta x_i
+ \ln_{2-q}\left(n\right)\ .
\end{equation}
Finally, taking the limit $n\rightarrow\infty$, we obtain the continuous form of the discrete nonadditive entropy as
\begin{equation}\label{q-entropy6}
S_{q}\left( \rho\right) = \lim_{n\to\infty}S_{q}(\{p\})= \int\limits_{a}^{b} \rho(x) \ln_{q}\rbr{\frac{m_q(x)}{\rho(x)}}\mathrm{d}x 
\,.
\end{equation}
where we omitted the divergent term $\lim_{n\to\infty}\ln_{2-q}(n)$ due to the same reason we omitted $\lim_{n\to\infty}\ln(n)$ in the Shannon case in Eq. (\ref{BGint2}). 
Namely, the physical observable is not the entropy itself but its change $\Delta S$, so that the divergence $\lim_{n\rightarrow\infty}\ln(n)$ in Eq. (\ref{BGint2}) and $\lim_{n\rightarrow\infty}\ln_{2-q}(n)$ in Eq. (\ref{q-entropy6}) for the Shannon and Tsallis entropy, respectively, vanishes, allowing the entropic structure to converge in the energy continuum.

Another issue worth noting is that the negative of the continuous expression $S\left( \rho\right)$ in Eq. (\ref{relent2}) for the BG entropy is nothing but the relative entropy (also known as Kullback-Leibler divergence) \cite{Jaynes,AbeBagci}, which reads
\begin{eqnarray}\label{KL1}
K\left[ \rho \| m\right]= \int\limits_{a}^{b}\rho (x) 
\ln \left( \frac{\rho(x) }{m(x) }\right)\mathrm{d}x   \,,
\end{eqnarray}
i.e., $-S\left( \rho\right) =  K\left[ \rho \| m\right]$.

Considering now the negative of the continuous nonadditive $q$-entropy in Eq. (\ref{q-entropy6}), we have
\begin{equation}\label{KL2}
-S_{q}\left( \rho\right) = - \int\limits_{a}^{b} \rho(x) \ln_{q}\rbr{\frac{m_q(x)}{\rho(x)}}\mathrm{d}x = \int\limits_{a}^{b} \rho(x) \ln_{2-q}\rbr{\frac{\rho(x)}{m_q(x)}}\mathrm{d}x
\,,
\end{equation} 
where we have used the relation $-\ln_q (x)=\ln_{2-q} (1/x)$ \cite{OikBagci2010}. The last expression above is exactly the Csisz{\'a}r-type nonadditive relative entropy $K_q \left[ \rho \| m\right]$ (see Eq. (24) in Ref. \cite{AbeBagci} or Ref. \cite{relentfirst} for example). In other words, just as its additive counterpart, i.e., $-S\left( \rho\right) =  K\left[ \rho \| m\right]$, the nonadditive entropy preserves the relation $-S_{q}\left( \rho\right) =  K_{q}\left[ \rho \| m\right]$ between its continuous generalization and the concomitant relative entropy expression.

So far we have shown that the term $n^{q-1}$ in Eq. (\ref{q-entropy1}) is essential, in the discrete case, to correctly obtain the concomitant continuous expression. The presence of this factor can further be elucidated by noting that the discrete entropy is maximized when the states are uniformly distributed. In other words, if we consider the discrete form of the relative entropy expression in Eq. (\ref{KL1}) with a uniformly distributed prior, i.e., $r_{i}=1/n$, then one obtains
\begin{eqnarray}\label{KL3}
K\left[ p  \| 1/n \right]
= \sum\limits_{i=1}^{n} p_{i} 
\ln \left( \frac{p_{i}  }{r_{i}  }\right) =\sum\limits_{i=1}^{n} p_{i} \ln \left( n p_{i} \right)
= -S(\{p\})-\ln(1/n)   \,
\end{eqnarray}
where $S(\{p\})$ denotes the discrete BG entropy in Eq. (\ref{BG}). The relation above shows that the entropy maximization is equivalent to the relative entropy minimization when the prior is chosen to be uniform \cite{Shore}. Therefore, the maximum entropy principle is a particular case of the relative entropy minimization.

A similar calculation using the discrete form of the nonadditive relative entropy $K_q \left[ p \| r\right]$ in Eq. (\ref{KL2}) with a uniform prior yields
\begin{eqnarray}\label{KL4}
K_{q} \left[ p  \| 1/n \right]
=- n^{q-1} \sum_{i=1}^{n}p_i \ln_q(1/p_i)-\ln_q\rbr{1/n}
=- S_q(\{p\})-\ln_q\rbr{1/n} \,
\end{eqnarray}
where the first expression on the right-hand side of the equality above is exactly the discrete entropy adopted in Eq. (\ref{q-entropy1}). In other words, the minimum relative entropy with a uniform prior is equivalent to the maximum discrete $q$-entropy expression $S_{q}(\{p\}) = n^{q-1} \sum_{i=1}^{n}p_i \ln_q(1/p_i)$, which explains the discrete form of the $q$-entropy adopted in Eq. (\ref{q-entropy1}) \cite{Jizba1,Jizba2}.

To conclude, we have shown that the discrete nonadditive $q$-entropy does indeed converge for any $q$ values. Moreover, the negative of the continuous $q$-entropy is shown to lead to the (Csisz{\'a}r-type) $q$-relative entropy mimicking exactly the relation between the negative of the continuous BG expression and the Kullback-Leibler relative entropy. Therefore, there is no obstacle for the use of the $q$-entropy to the continuous classical physical systems as many applications in the field also indicate \cite{Tsallisbook}.



\begin{thebibliography}{20}
\bibitem{Jizba1} Note that this same factor, i.e., $n^{q-1}$ is also noted in the third footnote of Ref. \cite{Jizba2} although there is no first principle derivation of the factor $n^{q-1}$ in Ref. \cite{Jizba2}.
\end{thebibliography}
\end{document}